# Sub-cycle modulation of light's Orbital Angular Momentum

Michael de Oliveira[1,2]*, and Antonio Ambrosio[1]*

[1]*Center for Nano Science and Technology, Fondazione Istituto Italiano di Tecnologia; Milano, Italy.*
[2]*Physics Department, Politecnico di Milano; Milano, Italy.*

*Corresponding authors: michael.almeida@iit.it; antonio.ambrosio@iit.it*

**The exploration of light has traditionally focused on its spatial properties, particularly its orbital angular momentum (OAM), while its temporal dynamics have remained an underexplored frontier due to the slow response times of existing modulation techniques. In this context, we introduce a method to modulate the OAM of light on a femtosecond scale by engineering a controllable space-time coupling in ultrashort pulses. By intricately linking azimuthal position with time, we implement a static, azimuthally varying wavefront transformation to dynamically alter the spatial distribution of light in a fixed transverse plane. Our experiments demonstrate self-torqued wave packets that exhibit spiraling motions and rapid temporal OAM changes down to a few femtoseconds. We further extend this concept to generate wave packets that angularly self-accelerate. We reveal that these wave packets dynamically adjust their OAM by redistributing their energy density across their spectral bandwidth, all without the influence of external forces. Owing to the unique properties of self-torque and angular acceleration, these time-varying OAM beams offer an accessible avenue for exploring light at fundamental time scales, with far-reaching implications for ultrafast spectroscopy, nano- and micro-structure manipulation, condensed matter physics, and other related areas.**

The orbital angular momentum (OAM) of light has consistently propelled the field of optical physics (*1*). Unlike spin angular momentum, which is directly linked to light's polarization, OAM arises from the spatial distribution of the light field, manifesting as a twisting of its phase front (*2*). This characteristic is quantitatively described by $e^{i\ell\phi}$, where $\ell$ represents the so-called topological charge, an integer that quantizes the OAM per photon in monochromatic beams, unveiling the inherently quantum mechanical facet of light (*3*). Recently, a novel frontier has emerged with the advent of 4D structured light, merging spatial complexities of OAM with temporal precision to create ultrafast space-time beams (*4*). These wave packets, characterized by their OAM, short pulse duration, and high peak intensity, have unveiled new spatiotemporal phenomena (*5*) including intertwined light coils (*6*, *7*), revolving-rotating beams (*8*) and toroidal pulses (*9*, *10*). On the other hand, other pioneering demonstrations have intentionally forsaken the collinearity between intrinsic OAM and linear momentum (*11–13*) to reveal phenomena like spatiotemporal vortices carrying transverse OAM (*14–17*).

Despite these strides, the invariant nature of OAM in these vortex pulses suggests an untapped potential, neglecting how these states can evolve or be controlled over time. The recent discovery of the self-torque of light, where OAM exhibits time-varying properties $\left(\text{i.e., } \xi = \frac{dL_z}{dt}\right)$, has revealed a new dimension of light's behavior, diverging from the static nature traditionally attributed to its angular momentum. This phenomenon was first observed in extreme-ultraviolet femtosecond vortex pulses, which arises in high harmonic generation excited by time-delayed vortex pulses with different topological charges (*18*). Unlike invariant OAM beams that exert a twisting force (i.e., torque) through angular momentum transfer during light-matter interactions, self-torque represents an intrinsic temporal evolution of OAM, leading to a self-induced rotation that alters the beam's phase and intensity profile. Remarkably, this bears resemblance to the kind of self-induced angular momentum variations observed in other domains (*19*), such as gravitational self-fields (*20*, *21*), offering a glimpse into the potential of dynamically controlling light's properties.

The primary challenge in effectively controlling the temporal dynamics of light for practical applications has been the slow modulation speeds of current technologies (*22–26*). A more promising avenue lies not in transient adjustments but in leveraging a singular static transformation based on the principles of wavefront shaping applied to a space-time coupled system. Notably, by correlating time and position through the speed of light ($z = ct$), it is possible to infer changes that occur over time (along its propagation path). This method allows for the creation of exotic beams following arbitrary trajectories (*27*), exhibiting angular acceleration in space (*28*), and changing wavelength and topological charge during propagation (*29*, *30*). However, while these achievements are impressive, they only reveal indirect temporal dynamics—at a fixed z-plane, these beams exhibit no time dependence and are limited by the length and time scales (typically centimeters and picoseconds) over which they operate.

This raises an intriguing possibility: if a unique form of space-time coupling could be engineered, one that operates independently of propagation, might we unlock the ability to directly manipulate the temporal dynamics of light? Achieving this would challenge existing paradigms and invite innovative approaches to modulate light with precision on a sub-cycle temporal scale. In this context, we propose and demonstrate an approach to control the temporal dynamics of light within a fixed transverse plane. Our approach involves engineering and exploiting a controllable space-time coupling that links azimuthal position with time. This is exemplified by the synthesis of a wave packet that uniquely spirals outward, characterized by a time-varying OAM beam and inherent self-torque. We further develop this concept by drawing a mechanical analogy, likening our beams to non-rigid bodies whose energy distributions are not necessarily stationary. In this way, we present a pioneering realization of angularly self-accelerating beams. These beams accelerate without the need for an external force, maintaining a constant radius and topological charge. The implications of this work are far-reaching, with potential applications in investigating particle collisions (*31*), selective excitation of magnetic (*32*), molecular (*33*) and quantum matter (*34*), to studying the optics of moving media (*35, 36*)—all within the ultrafast timescale of femtoseconds.

**Engineering space-time coupling in helical wave packets**

Our approach to tailor the temporal evolution of OAM within a fixed transverse plane involves engineering a unique space-time coupling mechanism—one that links the transverse profile with its temporal evolution and operates independently of propagation along the z-axis. One way to achieve this is by correlating azimuthal position ($\phi$) with time, which can be achieved by rotating a beam around a specific axis at an angular velocity, $\Omega$. The rotation introduces a space-time coupling expressed as $\phi(t) = \Omega t$.

Understanding the effects of this rotation involves examining the rotational Doppler effect. For instance, when a monochromatic vortex beam rotates with angular velocity $\Omega$, it undergoes a frequency shift that is measurable in the laboratory frame as $\omega_\ell = \omega' - \ell\Omega$ (*37*), where $\omega'$ represents the frequency in the rotating frame. Applying this to a rotating superposition of vortex modes—characterized by a common frequency $\omega'$, an average topological charge $\overline{\ell}$, and its spread $\Delta\ell$—results in a frequency spread $\Delta\omega = \Omega\,\Delta\ell$ in the laboratory frame, centred at the Doppler-shifted frequency $\omega_{\bar{\ell}}$ (see Supplementary Materials S3 for more details). Each vortex mode in the laboratory frame experiences its own frequency shift, establishing a relationship between frequency and topological charge according to:

$$\ell(\omega) = \bar{\ell} + \frac{\Delta\ell}{\Delta\omega}(\omega - \omega_{\bar{\ell}}). \tag{1}$$

This corresponds to the description of a helical wave packet (6), previously synthesized by the coherent superposition of vortex modes of different $\ell$, spread linearly across a spectrum of frequencies (see Fig. 1a). This combination manifests as a temporal beating, forming an envelope that follows a helical path on a cylinder in space-time. The spatial-temporal structure is elaborated in Fig. 1b and its caption (additional details in Supplementary Materials S2).

The practical implication is that we can engineer such a correlation by either physically rotating a monochromatic vortex superposition at an angular velocity $\Omega$, or alternatively synthesize the already Doppler-shifted frequency spread in laboratory frame. This suggests the presence of analogous mechanical properties, particularly angular momentum, in helical wave packets due to rotational motion. While both cases are equivalent manifestations in their respective reference frames, implementing the latter approach permits rotation speeds that far exceed the capabilities of current technologies. Experimentally, we generate the Doppler shifted wave packet, as defined by Eq. 1, directly in the laboratory frame. These helical wave packets serve as an intuitive basis from which we construct our temporally varying light fields, conceptually illustrates in Fig. 1c. For a detailed demonstration, refer to Supplementary Materials S4 and Movie S1, where we demonstrate a helical wave packet that reverses its orbital direction mid-trajectory.

In the following, we show that by strategically adjusting specific azimuthal segments of the wave packet, each defined by its own space-time correlation parameters ($\bar{\ell}(\phi)$ and $\Delta\ell(\phi)$), we can precisely tailor the wave packet's behavior across different time intervals. This results in different points along the azimuth to accrue different phase delays, causing the wave packet to spiral outwards (expanding in radius) or accelerate along its space-time path. Our innovative technique paves the way for unprecedented control over the temporal trajectory of light through a single static transformation, achieving sub-cycle temporal precision down to a few femtoseconds, significantly surpassing that of any time modulated device.

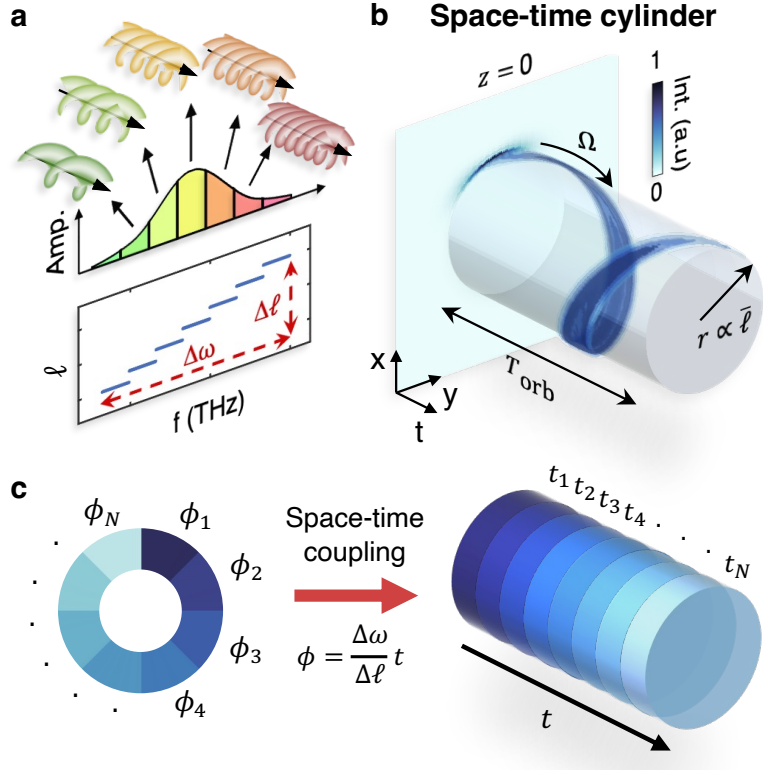

**Fig. 1. Space-time coupling in helical wave packets. a**) Helical wave packets are composed of a superposition of vortex modes correlated with a spectrum of frequencies of an ultrashort pulse. Their uniqueness is characterized by the mean topological charge $\bar{\ell}$, and its distribution $\Delta\ell$ across a frequency bandwidth, $\Delta\omega$. **b**) Simulated structure of a helical wave packet showing its transverse azimuthal profile, which arises from the superposition of $\ell$-modes. The linear correlation of these modes across frequency defines the wave packet's angular velocity, $\Omega = \Delta\omega/\Delta\ell$, thereby dictating its helical path confined to a unique cylinder in space-time. The mean topological charge $\bar{\ell}$ sets the radius of the cylinder, while its temporal extent is determined by the wave packet's angular velocity, $T_{\text{orb}} = 2\pi/\Omega$. **c**) At a fixed propagation plane ($z = 0$), the helical wave packet offers a direct coupling between azimuthal position and time, expressed as $\phi(t) = \Omega t$.

## Results

To generate helical wave packets, we implemented an all-digital Fourier space-time shaper (see fig. S4) consisting of two phase-only reflective spatial light modulators (SLMs) to manipulate our pulse. Our shaper is fed by a non-colinear parametric amplifier (NOPA) that emits NIR femtosecond pulses (760-840 nm) with a near transform-limited duration of 25 fs. The core of our setup is a digital axicon, a circularly symmetric diffractive grating, that disperses the bandwidth of frequencies along specific radial $k$-vectors to match the circular geometry of vortex beams. In the Fourier plane, the frequencies are radially separated, allowing us to apply independent phase transformations to each frequency, using a digital phase hologram. In addition to aberration correction and chromatic modulation, our holograms also introduce an innovative radial frequency grating to facilitate the separation in time of the desired wave packet from the zero-order component (see Supplementary Materials S1). To analyze and reconstruct the temporal evolution of these wave packets at a fixed propagation distance, we employ a combination of a delay line and off-axis holography. Further details can be found in Methods.

### *Time-varying Orbital Angular Momentum*

Until now, helical wave packets have been restricted to follow cylindrical trajectories in space-time, with their orbital radius associated with the wave packet's mean topological charge, $\bar{\ell}$, following a well-established relationship between radius r and charge $\ell$ in Bessel-Gaussian beams (*38*). In this study, we demonstrate that we can sculpt the space-time trajectory into a conical path, instead of cylindrical, with a radius that gradually increases over time (in a fixed transverse plane along the propagation axis), as depicted in Fig. 2a. This adjustment suggests a time-varying OAM density function, relying on the precise manipulation of the wavefront to change its number of helical twists over time. We achieve this by partitioning the momentum space of our wave packets into N equal azimuthal segments, each tailored with a different mean topological charge, according to $\bar{\ell}(\phi) = \bar{\ell} \pm \lfloor N\phi/2\pi \rfloor$, where $\lfloor \cdot \rfloor$ represents the floor function. The choice of N allows to adjust the variation in $\bar{\ell}(\phi)$, whereas its sign allows to create a vortex with linearly increasing (+) or decreasing (−) topological charge.

Our experimental realization involves segmenting the wave packet into $N = 50$ azimuthal divisions. Each segment maintains the same angular velocity $\Omega$ (and thus a fixed $\Delta\ell = 6$), but with an incremental increase in $\bar{\ell}$ from 101 to 150 (Fig. 2b). The resulting wave packet is illustrated in Fig. 2c (and Movie S2), clearly showing an increasing radius, a signature of increasing topological charge. This is further corroborated by the time-averaged transverse intensity profile spiraling outward in a linearly increasing manner analogous to an Archimedean spiral (Fig. 2d; see fig. S5 for additional characterization).

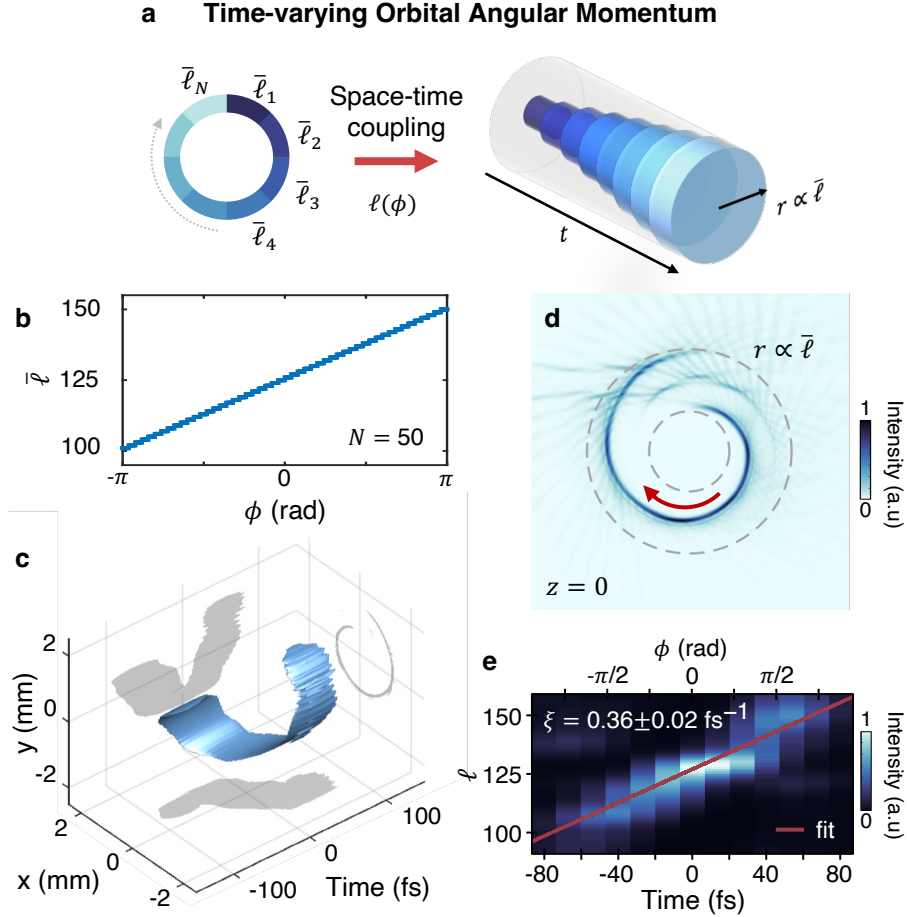

**Fig. 2. Time-varying orbital angular momentum. a**) By introducing an azimuthally varying dependence on the mean topological charge $\overline{\ell}(\phi)$, the space-time coupling allows us to sculpt the temporal trajectory of our wave packet towards a conical path, with a radius that increases over time. **b**) Segmenting the wave packet into $N = 50$ azimuthal divisions, each with a unique mean topological charge ranging from $\overline{\ell} = 101$ to $150$ and the same spread in $\Delta\ell = 6$. **c**) The corresponding experimentally reconstructed iso-surface depicting a wave packet tracing a spiraling conical path in time. **d**) The time-averaged transverse intensity profile of the wave packet, showcasing a spiraling pattern resembling an Archimedean spiral. **e**) Modal decomposition analysis confirming a continuous increase in $\overline{\ell}$, signifying a dynamically evolving OAM and a measured self-torque of $\xi = 0.36 \, \text{fs}^{-1}$. The iso-surface is represented at 15% peak intensity, and the time axes are normalized to center the wave packet at $t = 0$.

To verify the linear growth of mean topological charge $\overline{\ell}$, we perform a topological charge modal decomposition on the phase profiles retrieved through off-axis holography (see Methods). Our analysis, presented in Fig. 2e, confirms a monotonic temporal variation in $\overline{\ell}$, spanning 50 topological charges. In discussing the topological characteristics of our helical beams, it is imperative to emphasize a key distinction: the OAM density, $L_z$, of our helical beams is not simply quantized by the mean topological charge $\ell$, as is typical for monochromatic vortex beams ($11$, $12$). Instead, it includes a contribution due to the angular velocity of the wave packet. Therefore, the OAM of our field as derived from the change in wavefront as a function of azimuth (refer to Supplementary Materials S5), can be expressed as

$$L_z = -i \frac{d\varphi(\phi)}{d\phi} = \bar{\ell}(\phi) + \left(\frac{\Delta\omega}{\Delta\ell}t + \phi\right)\frac{d\bar{\ell}(\phi)}{d\phi}. \qquad (2)$$

This reveals a time dependence in the second term, resulting in a dynamically varying OAM. This contribution is observed in the rotating time-domain vortex due with an angular velocity, rather than the OAM of the frequency domain field which is a superposition of different vortex modes. Moreover, central to this is the conservation principle of OAM. While OAM is inherently quantized, it can exhibit local density variations that do not alter its overall density during its temporal evolution. More specifically, this is seen as different azimuthal positions accumulate different phase shifts in time, causing the helical twist of the wave packet to undergo continuous deformation and the emergence of phase singularities. As a result, our helical space-time beams contain a continuum of OAM states, arranged sequentially along its temporal trajectory.

We characterize the time-varying OAM spectrum of the wave packet using the concept of self-torque of light, defined as the rate of change of angular momentum, $\xi = \frac{dL_z}{dt}$. Simplifying the formulation by factoring out $\hbar$ and representing self-torque in units of fs$^{-1}$, we determine the self-torque of our beams to be described by:

$$\xi = \frac{\Delta\omega}{\Delta\ell}\frac{d\bar{\ell}(\phi)}{d\phi}. \qquad (3)$$

As a result, the self-torque can be attributed to the azimuthal change in mean topological charge, weighted by the angular velocity of the wave packet. This manifests as a change in the orbital radius of the light beam - akin to mechanical torque that arises from applying a wrench. This means that we can tailor the magnitude of the self-torque by either controlling the angular velocity (through $\Delta\ell$) or the azimuthal change in $\bar{\ell}$ of the wave packet. Analyzing our experimental results presented in Fig. 2d, a straight-line fit yields a self-torque of $\xi = 0.36\ \mathrm{fs}^{-1}$, implying a femtosecond variation of the OAM. This points to rapid sub-cycle OAM variations, where the OAM of the beam increases by approximately $\hbar$ every 2.7 femtoseconds. This OAM variation over time is significantly faster than the pulse duration, allowing us to differentiate our self-torqued beams from a sequence of non-overlapping pulses with different topological charges ($39$).

### *Angular self-acceleration of light*

In the next demonstration, we unveil a novel observation of the angular acceleration of light at a fixed $z$-plane, by realizing a wave packet that experiences an increasing angular velocity. Here, we discretize momentum space into $N = 8$ equal segments, but this time we azimuthally vary the angular velocity, $\Omega(\phi)$, while keeping $\overline{\ell} = 150$ constant (see Fig. 3a). Experimentally this is implemented by controlling the spread of topological charge according to $\Delta\ell(\phi) = \Delta\ell \pm \lfloor N\phi/2\pi \rfloor$ across a fixed frequency bandwidth $\Delta\omega$ (Fig. 3b, blue line). Since $\Delta\ell$ is inversely proportional to $\Omega$, a decreasing linear chirp in $\Delta\ell(\phi) = 13$ to 6, results in a non-constant acceleration, with an angular velocity from $\Omega = 0.018$ rad/fs to $0.039$ rad/fs (red line).

The dynamics of the generated wave packet are captured in Fig. 3c (and Movie S3), in which its temporal evolution reveals a speeding up of the wave packet. Figure 3d visually illustrates the angular position of the wave packet at three uniformly spaced time points on the time-averaged transverse intensity profile. It can be clearly seen that the wave packet traverses a larger angle from $t_2$ to $t_3$, compared to $t_1$ to $t_2$, indicating an angular acceleration. To precisely determine the parameters defining our wave packet's angular motion, we conduct an analysis of its angular position over time (see Fig. 3e), applying a third-degree polynomial fit (excluding negligible higher order terms). This allows us to extract the key parameters governing the wave packet's angular motion: of particular interest is the rate of change of acceleration, known as angular jerk, $j = -0.00001$ rad/fs³ demonstrating a non-constant acceleration, along with the initial parameters for angular velocity $\Omega_{t=0} = 0.04602$ rad/fs and angular acceleration $\alpha_{t=0} = -0.00015$ rad/fs².

As a result, our wave packet displays complex rotational dynamics due to its angular acceleration, a factor that significantly influences its behavior. These rotational dynamics must be considered to understand their impact on the wave packet's mean OAM density. Following the same derivation outlined as before (refer to Supplementary Materials S5), reveals a time-dependent contribution related to the azimuthal change in angular velocity:

$$L_z = \overline{\ell} + \overline{\ell} \frac{d}{d\phi} \frac{\Delta\omega}{\Delta\ell(\phi,t)} t. \tag{4}$$

Here, we note a contribution to the mean OAM despite a constant $\overline{\ell}$. This introduces a form of time-varying OAM that contrasts from previous demonstrations, which were primarily driven by changes in the mean topological charge $\overline{\ell}$, and consequently, the radius $r$. Indeed, in is clear from Fig. 3d that our wave packet maintains a constant orbital radius, and therefore a constant mean topological charge—instead, the variation in OAM results from the azimuthal chirp in its spread of topological charges $\Delta\ell(\phi)$, without altering the beam's path or radius.

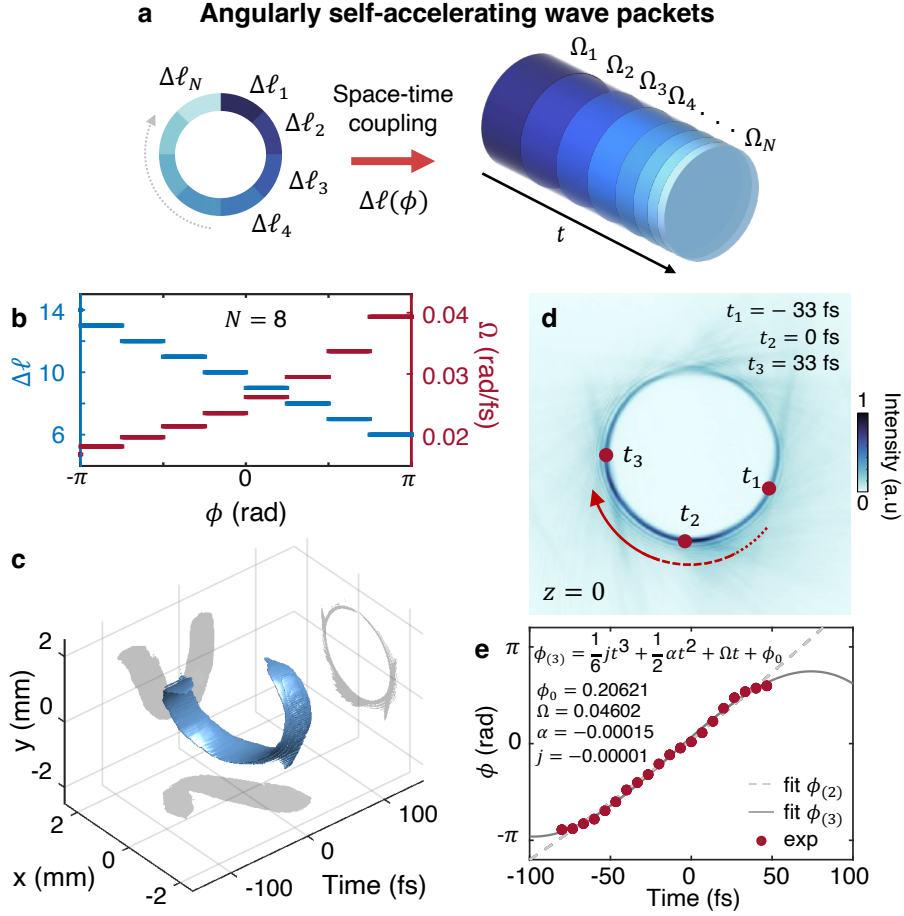

**Fig. 3. Angular self-acceleration of light without an external force. a)** Introducing an azimuthal chirp in $\Delta\ell(\phi)$ allows us to tailor the angular velocity $\Omega$ of the wave packet in time. **b)** The momentum space is partitioned into $N = 8$ segments, each with a constant mean topological charge $\bar{\ell} = 150$, but decreasing topological charge spread from $\Delta\ell = 13$ to 6 (blue). This results in an azimuthally increasing angular velocity (red), and a non-constant angular acceleration. **c)** The reconstructed iso-surface showcasing an increasing angular velocity. **d)** The time-averaged transverse intensity profile, confirming a constant orbital radius. We indicate the azimuthal positions of the wave packet at three equal time intervals, highlighting its accelerated motion. **e)** Analysis of the wave packet's angular position over time, fitted with a third-degree polynomial, providing insight into the wave packet's equations of motion. The iso-surfaces are represented at 15% peak intensity, and the time axes are normalized to center the wave packet at $t = 0$.

Similarly, we extend this understanding to the concept of the self-torque of light. The corresponding expression for this self-torque contribution, derived from the time derivative of the OAM density as before, is captured by:

$$\xi = \bar{\ell}\,\frac{d}{d\phi}\frac{\Delta\omega}{\Delta\ell(\phi,t)} + \bar{\ell}\,\frac{d}{dt}\frac{d}{d\phi}\frac{\Delta\omega}{\Delta\ell(\phi,t)}\,t\;. \tag{5}$$

This introduces an intriguing new perspective, suggesting a self-modulated angular acceleration that occurs without an external torque, much like a figure skater accelerating their rotation by drawing in their arms. In this context, angularly self-accelerating wave packets can be viewed analogously to non-rigid bodies in that their 'shape'—specifically, the distribution of topological charge across the frequency bandwidth, akin to mass distribution in a physical body—undergoes a change that influences their angular momentum and acceleration (see Supplementary Materials S6).

Moreover, prior studies have identified that a distinctive feature of self-torque is the presence of an azimuthal frequency chirp, where the frequency of the wave packet changes along its azimuthal angle (18). This characteristic becomes apparent in our beams when we analyze from the perspective of their rotating frame (see fig. S2). In our experiments, we engineer the frequency spread as it were shifted by the rotational Doppler effect, where the central frequency $\omega_{\bar{\ell}}$ and bandwidth $\Delta\omega$ are fixed by the initial pulse and remain unchanged in the laboratory frame. As a result, when viewed in its rotating frame, any changes in $\bar{\ell}(\phi)$ or $\Delta\ell(\phi)$ induce an effective angular frequency, displaying an azimuthal variation according to $\omega'(\phi) = \omega_{\bar{\ell}} + \bar{\ell}(\phi)\frac{\Delta\omega}{\Delta\ell(\phi)}$, thus supporting the presence of self-torque characteristics in our beams.

Our observations offer a profound understanding of self-torque, highlighting its correlation with the rotational dynamics induced by an azimuthal change in mean topological charge and/or angular velocity within our wave packets. This insight opens the possibility to tailor a dynamic, time-varying self-torque behavior. In fact, the non-constant angular acceleration, evident in our results in Fig. 3e, is already such a manifestation of time-varying self-torque, i.e., $\frac{d\xi}{dt} = \frac{d^2 L_z}{dt^2} \neq 0$. It further suggests that it is possible to harness higher-order OAM time derivatives for more sophisticated control, with our approach being readily adaptable to achieve any polynomial dependence in the azimuthally varying quantities $\bar{\ell}$ or $\Delta\ell$. Further details on how to optimize the range of achievable self-torque are discussed in Supplementary Materials S7.

## Discussion

Our results advance the experimental capabilities of shaping ultrashort wave packets with unrivaled control over their temporal and topological features, whilst also underscoring the potential for self-modulated dynamics in light beams. Though an innovatively tailored space-time coupling, we have unlocked the capability for sub-cycle temporal manipulation of light, in which we demonstrated the generation of time-varying OAM beams with intrinsic self-torque. Importantly, these rapid temporal scales open avenues for enhanced temporal precision in light-matter interactions involving OAM, including the potential for sub-cycle OAM-gating. Moreover, such spatiotemporal structures could arise naturally from moving sources that emit stationary (i.e., monochromatic) vortex states in their rest frames. Our beams could therefore serve as a temporal ruler to probe the rotational dynamics of a variety of systems involving moving frames or sources, across a broad spectrum of scales, from the quantum to the cosmic scale.

Our findings reveal a versatile generation mechanism that allows for custom shaping of the wave packet's spiral trajectory and angular acceleration over time. This flexibility offers a robust tool for optical trapping and manipulation, facilitating precise control over the motion of microscopic particles and nanostructures. Furthermore, by modifying the function that discretizes the momentum space, it's possible to generate beams with higher-order time derivatives of OAM. Extending spectral control to adjust amplitude and phase among topological-spectral modes could lead to even more complex light structures and dynamics, such as an orbiting donut vortex (*8*). While our current focus is on scalar waves, integrating our method with spin-orbit optics like metasurfaces (*40*, *41*) opens the door to exploring the temporal dynamics of vector waves carrying spin angular momentum. This integration could yield exotic polarization structures with fascinating, time-varying polarization states, offering a new layer of complexity in light manipulation.

All of this can be used to deliver optical torque within the natural time and length scales of charge and spin ordering, opening exciting prospects in ultrafast spectroscopies of angular momentum dynamics, laser-plasma acceleration, as well as imaging and manipulation of molecules and nanostructures on their intrinsic scales. Additionally, our research suggests potential parallels in other natural systems beyond light, extending to wave structuring in acoustics (*42*, *43*), electrons (*44*, *45*) and quantum matter (*34*, *46*). Ultimately, the ability to precisely control light at such a fundamental level heralds new frontiers for scientific and technological advancements.

## Methods

### *Digital Fourier space-time beam shaper*

Our space-time beam shaper, designed to generate helical wave packet, employs an approach inspired by conventional pulse shapers (*47*). Instead of a traditional linear grating, we use a digital axicon grating applied via a spatial light modulator (SLM) to disperse frequencies into concentric rings. A non-collinear optical parametric amplifier (ORPHEUS-N) serves as the coherent light source, emitting NIR femtosecond pulses (760-840 nm) with near transform-limited duration of 25 fs (see fig. S4a). The NOPA beam, with a beam waist of $w_0 = 3$ cm, is aligned into our beam shaper (see fig. S4b), with its polarization adjusted using a broadband half-wave plate to match the modulation axis of the SLMs. The beam is directed to the first SLM (HOLOEYE PLUTO-2.1-NIR-133 module, 8 μm pixel pitch), which imparts an axicon grating function through a radially symmetric phase hologram. By digitally tuning the axicon's grating period and size, we optimize the spectral resolution of our space-time shaper. A $2\pi$ phase shift is calibrated for a central wavelength of 800 nm. We set the grating period to $d = 32$ μm, using a minimum of 4 pixels impart a phase ramp from 0 to $2\pi$. A broadband fused silica beam splitter (BSW11R, 1 mm thick) directs the beam at normal incidence to the SLM and creates the interferometer for holographic measurements.

We use a spherical concave mirror ($f = 150$ mm) to perform a broadband spatial Fourier transform, providing direct access to the far-field of the axicon. At this Fourier plane, the pulse appears as concentric rings with wavelengths distributed linearly along the radial direction. The radius and half-thickness of these rings are defined by $R = f\lambda/d$ and $T = f\lambda/\pi w_0$, respectively (*48*). We use a monochromatic 800 nm laser source (SuperK Select) and observe a spectral resolution ($\delta\lambda = \lambda 2T/R$) between 4 and 5 nm across the pulse bandwidth. This indicates that our system can address up to 16 independent spectral windows across the spectral bandwidth of our pulse. Since we are only concerned with applying a linear function in $\ell(\omega)$ that transverses $\ell$ in steps of 1, this sets the maximum range in $\Delta\ell$ that we can achieve. We show that even with only 7 spectral windows (i.e., $\Delta\ell = 6$), we can generate helical wave packets, reducing the need for ultra-short femtosecond pulses or very large bandwidths, thereby enhancing the accessibility and versatility of our approach.

A second high-resolution 4K SLM (HOLOEYE GAEA-2 NIR module, 3.74 μm pixel pitch) at the Fourier plane imposes the spectrally dependent spatial structure of our beams. The displayed phase hologram is segmented into a series of concentric rings, each ring imparting a unique topological charge to a different spectral window. The hologram also corrects for inherent astigmatic aberrations using Zernike polynomials. Additionally, we superimpose a radial frequency grating to mitigate zero-order artifacts that arise from modulation inefficiencies in typical setups (refer to Supplementary Materials S1 for details). A key step involves calibrating the wavelength-specific response of the SLM across the pulse's spectral range, allowing us to map the desired phase profile to precise SLM voltage levels. We achieve this broadband response with the SLM because the wavelengths are spatially dispersed at the second SLM plane. After the second SLM, a concave mirror ($f = 200$ mm) performs an inverse Fourier transformation to generates our helical wave packets at the image plane of the first SLM.

### *Off-axis holography*

To characterize the time-domain evolution of our beams at a fixed propagation distance, we use a spatial interferometer based on off-axis holography. Experimentally, we extract the unstructured reference pulse using the first beamsplitter in our setup, which we then interfere at an angle with our structured helical beam on another beamsplitter. Since both beams are pulsed, only their temporally overlapping portions contribute to the interference signal, controlled by adjusting the delay in the setup. The delay line, consisting of a hollow roof mirror mounted on a motorized micrometer stage, precisely adjusts the reference pulse path length and scans through various time delays. The resulting interference pattern is magnified with concave mirrors and imaged using a high-resolution camera (DMK 38UX253, The Imaging Source).

The wave packets are imaged at a propagation distance of z = 0 cm relative to the imaging plane ($4f$) of the first SLM. Background measurements are also taken and subtracted from the measured interference pattern to reduce the d.c. components in the digital Fourier transform. For each time delay image, we apply the procedure of off-axis digital holography (*49*): the interference pattern is digitally Fourier transformed, and one of the first order terms is isolated with an annular ring. Applying an inverse Fourier transform provides the amplitude and phase profiles of our wave packets relative to the reference beam. Systematically scanning the time delay, we reconstruct the evolution of the beam's structure over time, represented as an iso-surface of 15% of the peak intensity. We note that the qualitative features of the iso-surface remain independent of the specific threshold value chosen.

### *Modal decomposition*

To measure the dynamic topological charge of a wave packet over time, we employed a digital modal decomposition approach. This method involves analyzing the reconstructed complex field from each time-delay image, performing an overlap integral with phase-only vortex eigenstates to extract their weightings (*50*). These eigenstates, while not forming an orthonormal basis, allow to assess the topological charge content within the optical field. It is important to note that these measurements are performed in the time domain of our helical wave packets and give access to the instantaneous topological charge distribution at each time slice, rather than exploring its distribution in the frequency domain.

**Acknowledgments:** We thank A. Petrozza, EL. Wong and G. Folpini for generously providing access to and training on the NOPA. This work has been financially supported by the European Research Council (ERC) under the European Union's Horizon 2020 research and innovation programme 'METAmorphoses', grant agreement no. 817794.

# Supplementary Materials to:
# Sub-cycle modulation of light's Orbital Angular Momentum

Michael de Oliveira[1,2]*, and Antonio Ambrosio[1]*

[1]Center for Nano Science and Technology, Fondazione Istituto Italiano di Tecnologia; Milan, Italy.
[2]Physics Department, Politecnico di Milano; Milan, Italy.

*Corresponding authors: michael.almeida@iit.it; antonio.ambrosio@iit.it

**The PDF file includes:**

    Supplementary Text
    Figs. S1 to S6

**Other Supplementary Materials for this manuscript include the following:**

    Movies S1 to S3

## S1. Radial frequency grating for order separation in time

Wavefront shaping techniques, particularly those employing SLMs, are inherently lossy, achieving conversion efficiencies typically in the range of 60-80%. The primary difficulty lies in isolating the desired optical field from the background of unmodulated components. Traditionally, this has been addressed by directing light into specific diffraction orders and then employing spatial linear gratings to angularly separate different diffraction orders in space or using polarization as a drop-port via metasurfaces (*1*). In our case, our phase-only SLMs do not modulate polarization and using a spatial grating is not suitable for our broadband source, which would make it nearly impossible to preserve the collinearity of our frequencies.

In this context, we innovated by leveraging time as a drop-port, opting to separate the diffraction orders in the time domain and preserve their collinearity. We achieve this by employing an innovative frequency grating, implemented as a radial phase gradient across the radially dispersed frequencies in the Fourier plane—what we call a frequency axicon. This is mathematically represented as

$$\varphi_g\left(\omega\right) = \frac{2\pi f}{\Lambda} = \frac{\omega}{\Lambda}, \tag{S1}$$

where $\Lambda$ is the frequency grating period. Analogous to a spatial axicon that induces a phase transformation that is radial and linear relative to the distance from the optical axis, our frequency axicon uniquely applies a phase transformation that linearly varies with the radially dispersed frequency rather than spatial position. An example of the hologram used to generate the helical wave packets, incorporating the topological correlation and frequency grating, is shown in fig. S1a.

A key outcome of this frequency-dependent phase is a temporal shift within the time-domain. More specifically, the wave packet experiences a group delay, mathematically represented as the derivative of its phase with respect to the angular frequency:

$$\tau_g = -\frac{d\varphi_g(\omega)}{d\omega} = -\frac{1}{\Lambda}. \tag{S2}$$

In physical terms, the group delay quantifies how the peak, or the envelope, of the wave packet is delayed by the system. This term arises from the grating applied across the frequencies, imposing a constant delay across the entire envelope, and shifting it in time. Drawing an analogy with spatial gratings, this creates a diffraction order in time at $\tau_g$.

For all our experiments, we standardize the frequency grating period at $\Lambda = 5.6$ THz. Figure S2b shows the helical wave packet experimentally generated using the hologram in fig. S1a, measured at a fixed propagation distance of z = 0. We clearly see that the helical wave packet is temporally separated from the zero-order component. The group delay of the generated helical envelope was measured as the time when the wave packet is aligned with the azimuthal position $\pi$, marking the helical wave packet's center (dashed red line in fig. S1b). The group delay was measured to be $\tau_g = -173$ fs, showing good agreement with theoretical prediction. This manipulation enables precise control over the temporal positioning of the wave packet, effectively separating the zero-order component which often represents an undesired artifact arising from modulation inefficiencies in typical setups (see the intensity profiles in fig. S1c). We normalize all the generated wave packets in the paper by adjusting for this group delay, thereby centering our wave packets at $t = 0$.

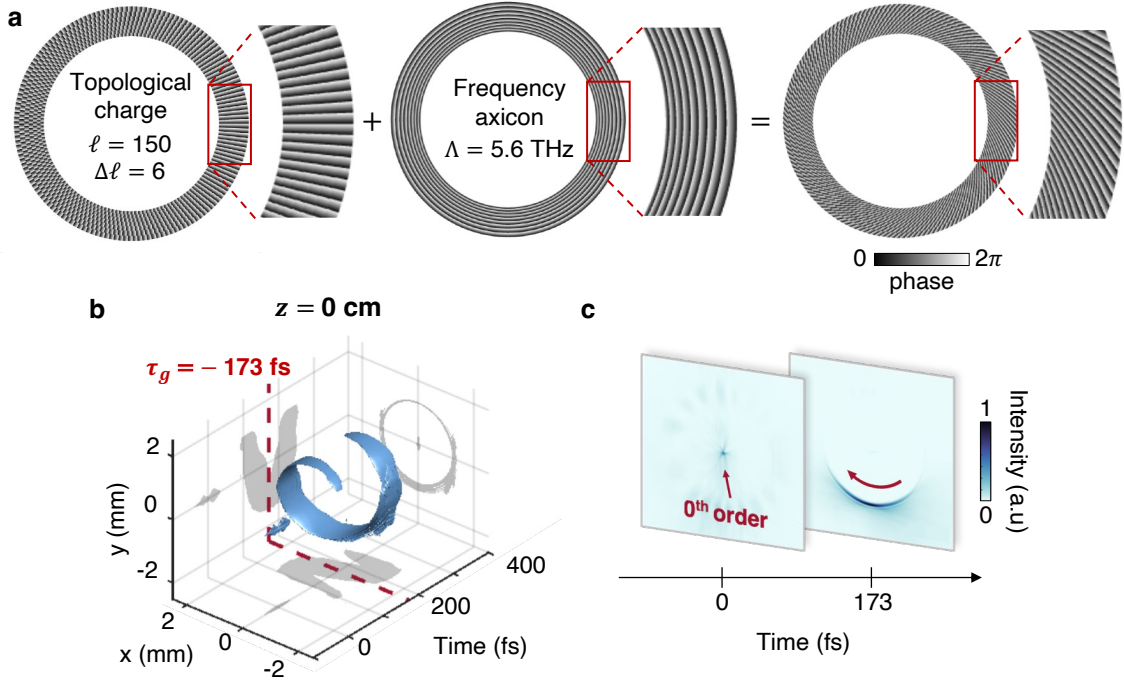

**Fig. S1. Radial frequency grating for order separation in time. a**) An example of a digital phase hologram (right), which assimilates the vortex phase of the topological spectral correlation (left) with a radial frequency grating (center). **b**) The reconstructed iso-surface (15% peak intensity) of helical wave packets (measured at a propagation distance, $z = 0$), generated with a frequency grating period $\Lambda = 5.6$ THz. The frequency grating imparts a group delay on the envelope of the wave packet, shifting it in time and separating the wave packet from the unstructured zero order. The dashed red line indicates the center of the helical structure, which is determined where the wave packet aligns with the azimuthal position $\pi$. **c**) The corresponding intensity profile of wave packet's complex electric field obtained via off-axis digital holography with a delay line.

## S2. Inherent space-time coupling in helical wave packets

Consider a helical wave packet where each vortex mode is associated with a different frequency. Such a wave packet is characterized by a mean topological charge $\overline{\ell}$ associated with the central frequency $\omega_{\overline{\ell}}$, and exhibits a spread $\Delta\ell$ across the frequency range $\Delta\omega$ as defined by Eq. 1 of the main text. The electric field describing this helical wave packet is expressed as a superposition:

$$E(r, \phi, z, t) = \sum_{\ell} |BG_{\ell}| \exp[i(-\omega t + \ell(\omega)\phi + kz)], \tag{S3}$$

where $r$, $\phi$, $z$ and t denote the radial, azimuthal, propagation and time coordinates respectively, and $k = \frac{\omega}{c}$. The transverse profile of the beam, described by $|BG_\ell| = A_\ell \exp\left(-\frac{r^2}{w^2}\right) J_\ell(k_r r)$ is shaped by higher-order Bessel-Gaussian modes, where $J_\ell$ represents the Bessel function of the first kind. The spatial distribution of the vortex modes interferes to form an azimuthal profile. The frequency-dependent spread of $\ell(\omega)$ induces a linear variation in relative phases over time, manifesting as an azimuthal group delay. Consequently, the azimuthal position of the profile rotates over time such that $\phi \propto t$.

This dynamic interplay reveals the wave packet's spiraling trajectory, whereby the intensity maxima occur when the phase $\varphi = -\omega t + \ell(\omega)\phi + kz$ is stationary with respect to $\omega$, satisfying the condition $\frac{d\varphi}{d\omega} = 0$. This leads to the spatial and temporal locations $\phi(t, z)$ of the pulse maxima:

$$\phi(t, z) = \frac{\Delta\omega}{\Delta\ell}\left(t - \frac{z}{c}\right), \tag{S4}$$

where $\Omega = \frac{\Delta\omega}{\Delta\ell}$ is the angular velocity, leading to an orbital period $T_{\text{orb}} = \frac{2\pi}{\Omega}$ at any given $z$. By rearranging eq. S4, we find that $t = \frac{z}{c} + \frac{\phi}{\Omega}$. The restructured equation enables us to delineate a space-time coupling that, in addition to the inherent correlation between propagation and time via the speed of light $c$, also depends on the angular velocity $\Omega$ of our wave packet. At a fixed transverse plane ($z = 0$), this space-time coupling reduces to a direct correlation between azimuthal position and time, expressed as $\phi = \Omega t$. This concept is central to our work, enabling us to induce temporal changes via an azimuthally varying modulation within a fixed $z$-plane.

## S3. Rotational dynamics of helical wave packets

To further deepen our understanding of the intricate spatial-temporal structure of helical wave packets, it is beneficial to view them as a rotating body within a fixed transverse plane (such as at $z = 0$). In this way we can glean valuable insights about their rotation dynamics. In the following, we will examine a helical wave packet as a monochromatic superposition of vortex modes that are set into rotation. This is intuitively illustrated and summarized in fig. S2.

A key phenomenon in our discussion is the rotational Doppler effect (*2*), in which a monochromatic vortex beam, characterized by a frequency $\omega'$ and topological charge $\ell$, experiences a Doppler frequency shift when rotated with an angular velocity $\Omega$. Here, the primed variable indicates the rotating reference frame. In the laboratory frame of reference, this shift adjusts the frequency of the vortex mode to $\omega = \omega' - \ell\Omega$ (illustrated in fig. S2a), which depends both on the value of $\ell$ and $\Omega$.

We can further extend this concept to a field composed of a monochromatic superposition of vortex modes of different $\ell$, each at the same frequency $\omega'$, as shown in Fig. S2b (left column). The stationary, non-rotating field is described by:

$$E(r, \phi, 0, t) = \sum_\ell |BG_\ell| \exp[i(-\omega' t + \ell\phi)], \tag{S5}$$

where each mode interferes to create an azimuthal profile localized around a fixed angle $\phi$. We then set the field into rotation with angular velocity $\Omega$. The rotation can be mathematically represented by a time-dependent coordinate rotation operator, which in cylindrical coordinates transforms a function according to $T_{\Omega t}[f(r, \phi)] = f(r, \phi + \Omega t)$. This is equivalent to giving each component a time-dependent factor $\exp(i\ell\Omega t)$, so that the rotating profile has the form:

$$
\begin{aligned}
T_{\Omega t}\left[E(r, \phi, 0, t)\right] &= \sum_{\ell}|BG_{\ell}| \exp\left[i\left(-\omega' t + \ell(\phi + \Omega t)\right)\right], \\
&= \sum_{\ell}|BG_{\ell}| \exp\left[i\left((-\omega' + \ell\Omega)t + \ell\phi\right)\right].
\end{aligned} \tag{S6}
$$

In the laboratory frame, the rotation induces a frequency shift in each vortex mode, proportional to its $\ell$ value, expressed as $\omega_{\ell} = \omega' - \ell\Omega$. This shift follows from a natural manifestation of the rotational Doppler effect and results in a spread of frequencies $\Delta\omega = \Omega \Delta\ell$ centered at $\omega_{\bar{\ell}}$. The rotating beam can therefore be described as a superposition of the vortex modes with discrete, equally spaced frequencies, as shown in fig. S2b (right column).

Upon comparing eq. S6 with eq. S4 at a fixed $z$-plane (i.e., $z = 0$), it becomes clear that in the rotating frame, a helical wave packet is observed as a monochromatic superposition of vortex beams (of a single frequency, $\omega' = \omega_{\ell} + \ell\Omega$) that rotates around the origin. This perspective offers a unique framework to examine the profound effects and inherent behaviors of rotational dynamics within the optical domain (3). In practice, this means that we can either physically initiate the rotation of a monochromatic superposition of vortex modes or alternatively engineer the already Doppler shifted frequency spread in the laboratory frame. While both outputs are theoretically equivalent in their respective reference frames, implementing the latter approach permits rotation speeds that far exceed the capabilities of current mechanical, electrical, or digital means. Experimentally, our approach generates the Doppler shifted wave packet directly in the laboratory frame—this being the equivalent of physically rotating a beam with an angular velocity in the terahertz range.

We can now use this framework to help glean further insights about the rotational dynamics of self-torque wave packets, like those demonstrated in Fig. 2 and 3 of the main text. The two cases, either changing $\bar{\ell}(\phi)$ or $\Delta\ell(\phi)$, are illustrated in fig. S2c and S2d, respectively. In our experiments, the central frequency $\omega_{\bar{\ell}}$ and the frequency spread $\Delta\omega$ of the wave packets are predefined by the characteristics of the input pulse and remain constant regardless of the applied correlation. However, when viewed from a rotating frame, the effective angular frequency, $\omega'$, exhibits an azimuthal variation according to $\omega'(\phi) = \omega_{\bar{\ell}} + \bar{\ell}(\phi) \frac{\Delta\omega}{\Delta\ell(\phi)}$, since $\omega_{\bar{\ell}}$ is constant but $\bar{\ell}(\phi)$ and $\Delta\ell(\phi)$ are azimuthally varying. The resulting azimuthal frequency chirp—where the frequency of the wave packet changes along its azimuthal angle—is a distinctive feature of self-torque beams, as reported in the literature.

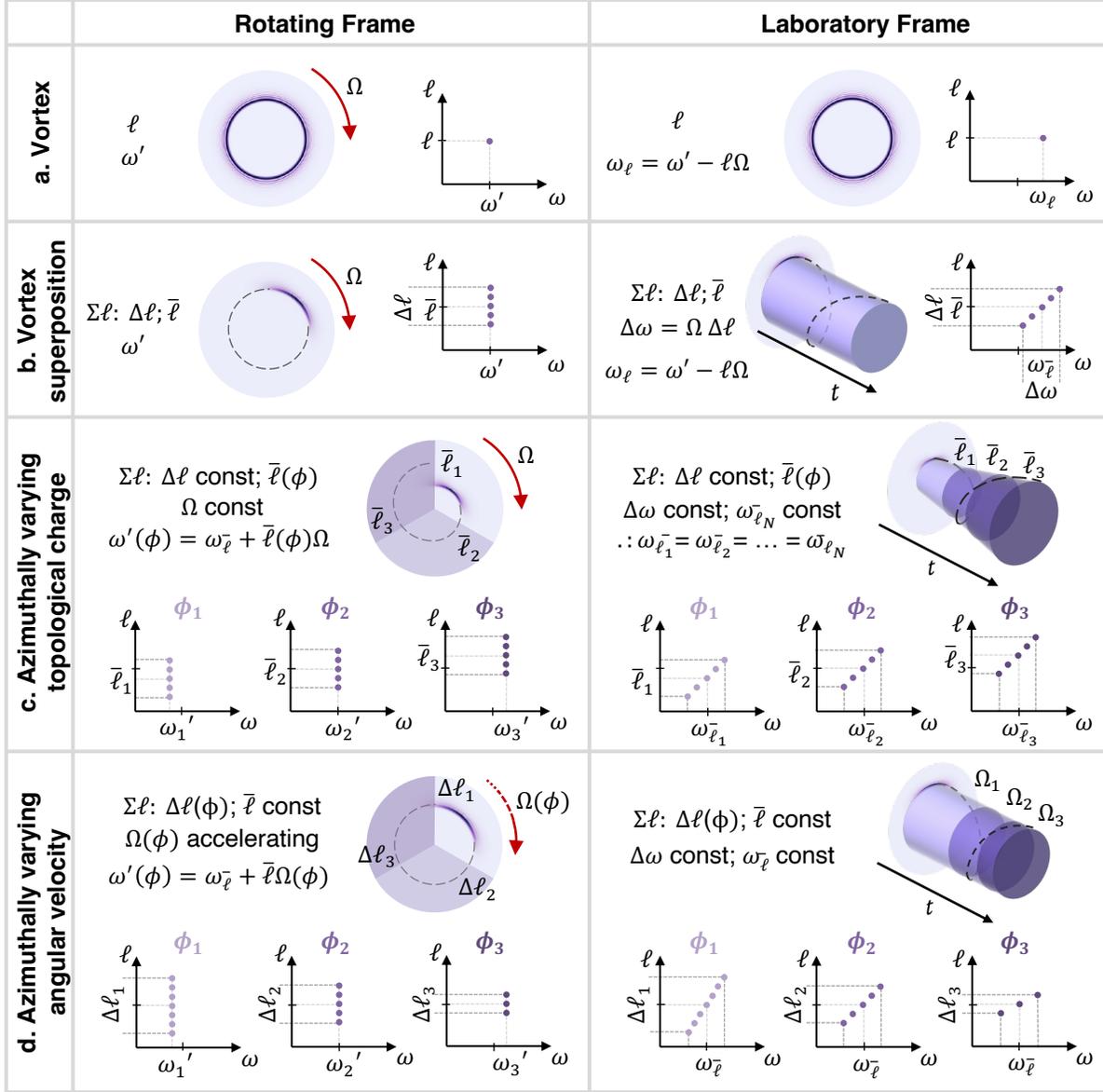

**Fig. S2. Rotational Doppler effect in helical wave packets.** Conceptual illustration of the dynamics observed in a rotating frame (left column) and the laboratory frame (right column). **a**) A monochromatic vortex beam with a frequency $\omega'$ and charge $\ell$. When this beam is rotated at an angular velocity $\Omega$, it undergoes a Doppler frequency shift to $\omega_\ell = \omega' - \ell\Omega$ in the laboratory frame. **b**) A superposition of vortex modes, each at the same frequency $\omega'$, creating an azimuthal profile through their interference. Upon rotation, each mode experiences a Doppler shift, resulting in a spread of frequencies $\Delta\omega$ centered at $\omega_{\bar{\ell}}$. This forms a helical wave packet equivalent to the wave packets we generate experimentally. **c**) and **d**) further extend this concept to self-torque wave packets, where either $\ell(\phi)$ or $\Delta\ell(\phi)$ is varied azimuthally (different shaded regions). In both scenarios, these beams are produced in the lab frame where the central frequency $\omega_{\bar{\ell}}$ and spread $\Delta\omega$ are fixed by the input pulse. However, from a rotating frame, the angular frequency $\omega'$ shows an azimuthal variation following $\omega'(\phi) = \omega_{\bar{\ell}} + \bar{\ell}(\phi)\frac{\Delta\omega}{\Delta\ell(\phi)}$.

## S4. Tailoring the temporal trajectory of light

Here, we experimentally demonstrate the operating principle underlying our technique, which leverages the inherent space-time coupling in helical wave packets. Central to our method is the Fourier relationship between angular position and OAM (*4*). Using a Fourier space-time shaper grants us precise access to the momentum space of our wave packet. As illustrated in fig. S3a, selecting a specific segment in momentum space (highlighted in purple or red) corresponds to isolating specific portions of the wave packet, as shown in fig. S3b. The extent of the temporal localization depends on the size of the selected wedge in momentum-space. Applying an additional constant phase ($\pi$) across the frequencies allows us to rotate the wave packet (180°), altering its start and end points (fig. S3c). Now, selecting the same azimuthal segments in momentum space leads to portions of the wave packet localized around the same azimuthal position but occurring at different times compared to those shown in fig. S3b—in this case even reversing their temporal order. This enables us to construct azimuthal segments of a wave packet and meticulously assemble them, providing precise control over their temporal positioning.

As an initial demonstration of this approach, we successfully generate a single helical wave packet that reverses its orbital direction midway through its temporal trajectory. This phenomenon represents a significant departure from the conventional behavior of helical wave packets previously demonstrated, which maintain a consistent orbital direction as they move along their trajectory on the space-time cylinder.

We divide the momentum space into two segments and apply conjugate topological-spectral correlations to each segment (fig. S3d). One segment applies a correlation function with $\bar{\ell} = 150$ and $\Delta\ell = 6$ (left), generating an azimuthal wave packet that follows a clockwise orbit and completing only half a rotation. Conversely, we apply the conjugate correlation function with $\bar{\ell} = -150$ and $\Delta\ell = -6$ (right) to the other segment. This reverses the rotation direction of our wave packet, completing a half rotation in an anticlockwise direction. We also add a constant $\pi$ phase across the frequencies in the second segment to shift in time and align the start point of the second azimuthal wave packet to the end of the first azimuthal wave packet. By piecing together both semi-arch phase transformations, we succeed in generating a single wave packet that reverses its chirality. The corresponding iso-intensity surfaces and intensity profile evolutions are presented in fig. S3e and S3f, respectively. See also Movie S1.

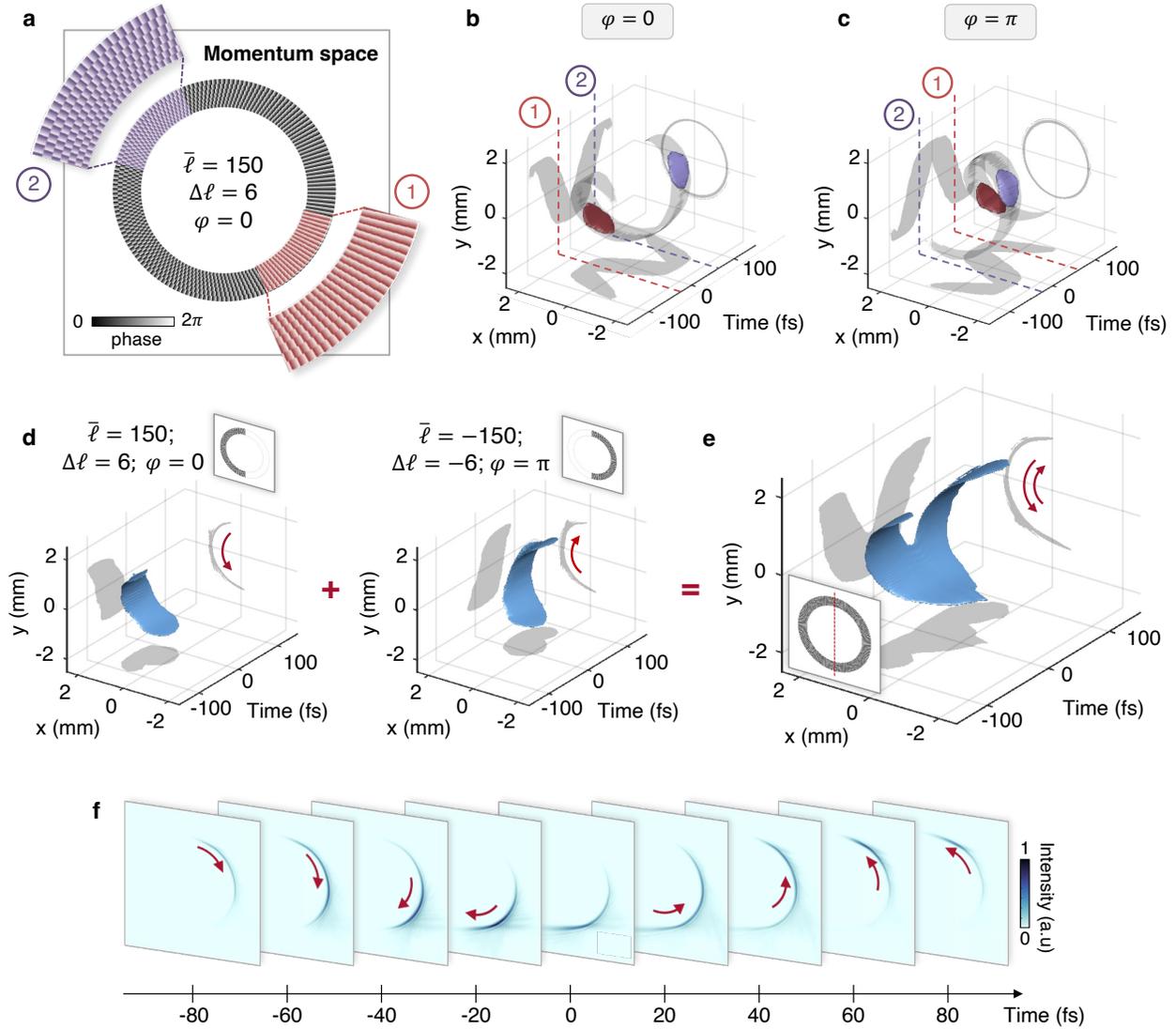

**Fig. S3. Crafting and reversing helical wave packet trajectories. a)** Schematic representation of the momentum space of our wave packet. Selecting specific azimuthal portions in the wave packet's momentum space (highlighted in purple or red segments), results in temporally localized portions of the wave packet as shown in **b**). **c)** Adding a $\pi$ phase across the frequencies rotates the wave packet by 180°, leading to portions of the wave packet localized about the same azimuthal positions but at different times compared to those in **b**). **d)** Experimental demonstration of two halves of a helical wave packet, achieved by dividing the momentum space into two segments and applying conjugate topological-spectral correlations: (left) $\bar{\ell} = 150, \Delta\ell = 6, \varphi = 0$ and (right) $\bar{\ell} = -150, \Delta\ell = -6, \varphi = \pi$. A constant $\pi$ phase is added to the latter, to align the start and end points of the two wave packets. **e)** Combining these two segments successfully generates a single wave packet that notably reverses its rotational direction midway through its trajectory. The iso-surfaces are shown at 15% peak intensity. **f)** Intensity images of wave packet's complex electric field (from **c**)) obtained via off-axis digital holography.

## S5. The OAM of space-time helical wave packets

For monochromatic vortex beams, characterized by a helical twist in their wavefront, it is well stablished that the intrinsic OAM is colinear with the beam's linear momentum, and is quantized by the helicity or topological charge $\ell$. However, this relationship becomes more complex when extended to polychromatic beams (5, 6). In this context, we extend the notion of OAM density to encompass polychromatic states, notably those where the topological charge varies with frequency. Through this expansion, we formulate an equation to quantify the mean intrinsic OAM of helical wave packets, showing that the OAM density is influenced by their inherent angular velocity and acceleration which these beams can exhibit.

Let us begin by considering an arbitrary mode profile that is rotating with a constant angular velocity in the transverse plane $z = 0$, according to eq. S5. Fully elucidating the angular dynamics of such a helical wave packet requires consideration of both time- and space-averaged energy characteristics of the electromagnetic field, which is complex. Instead, we aim to provide some insights into these dynamics by focusing on the mean OAM density, reducing the sum of vortex modes to a single vortex beam characterized by the average topological charge $\bar{\ell}$ that is rotating with angular velocity $\Omega$. We can proceed to describe the mean OAM carried by a helical wave packet, which can be quantified by the rate of change of the wavefront's phase $\varphi(\phi, t)$ with respect to the azimuthal coordinate. Mathematically, this is expressed as:

$$L_z = -i \frac{d\varphi(\phi,t)}{d\phi}, \tag{S7}$$

In the case of a helical wave packet, the mean OAM density is simply given by, $L_z = \frac{d}{d\phi}(\omega' t + \bar{\ell}\phi + \bar{\ell}\Omega t) = \bar{\ell}$ — a rather trivial result.

Now consider that we can impose an azimuthal dependency on the two parameters $\bar{\ell}(\phi)$ and/or $\Delta\ell(\phi)$, with the latter corresponding to an azimuthal change in the angular velocity $\Omega(\phi, t)$. According to eq. S7, the mean instantaneous OAM density carried by the beam is then described as:

$$\begin{aligned} L_z &= \frac{d}{d\phi}\left(\omega' t + \bar{\ell}(\phi)\phi + \bar{\ell}(\phi)\Omega(\phi,t)t\right) \\ &= \bar{\ell}(\phi) + (\Omega(\phi,t)t + \phi)\frac{d\bar{\ell}(\phi)}{d\phi} + \bar{\ell}(\phi)\,\frac{d\Omega(\phi,t)}{d\phi}t. \end{aligned} \tag{S8}$$

The equation above can be understood in three distinct parts: a base term and two time-dependent components. The first term represents the conventional understanding of OAM, quantized by the topological charge $\bar{\ell}$. When there is no azimuthal variation in $\bar{\ell}$ or $\Delta\ell$, this term alone describes the intrinsic OAM as traditionally understood. The second term encapsulates the changes in OAM attributable to the azimuthal variation in the topological charge $\bar{\ell}(\phi)$, which is akin to the contribution to time-varying OAM that has been previously demonstrated in literature (7–10). The third term is particularly interesting; it introduces a novel aspect that was previously not considered or demonstrated in the context of electromagnetic waves. It accounts for changes in the angular velocity of the beam, where a change in $\Delta\ell(\phi)$ can be translated as an effect of wave packet's angular acceleration.

These terms manifests as local OAM density variations that do not alter the beam's global density throughout its evolution, ensuring the inherent conservation of OAM. Specifically, when changes occur in $\bar{\ell}(\phi)$ or $\Delta\ell(\phi)$, the OAM is conserved either through adjustments in the beam's radius or through the redistribution of its energy density across its spectrum, respectively. Consequently, the dynamic and evolving wavefront structure of such azimuthally varying helical wave packets contain all intermediate OAM states, which are sequentially arranged in time along the pulse.

## S6. The implications of the self-torque of light

The concept of self-torque in light beams is defined as the intrinsic change in OAM over time, described by $\xi = \frac{dL_z}{dt}$. Applying this to the mean OAM density of our beam, the self-torque phenomenon in our beams is given by the equation:

$$\xi = \frac{d\bar{\ell}(\phi)}{d\phi}\Omega(\phi, t) + \bar{\ell}(\phi)\frac{d\Omega(\phi,t)}{d\phi} + \bar{\ell}(\phi)\frac{d}{dt}\left(\frac{d\Omega(\phi,t)}{d\phi}\right)t. \tag{S9}$$

This equation decomposes into three key terms:

The first term encapsulates a self-torque dependent on the product of the angular velocity of the wave packet and the azimuthal change in mean topological charge, $\bar{\ell}(\phi)$. Under a constant angular velocity, the angular position exhibits a linear time dependency, $\phi = \frac{\Delta\omega}{\Delta\ell}t$, allowing us to reformulate the self-torque equivalently as a rate of change of $\bar{\ell}$ over time, i.e., $\xi = \frac{d\bar{\ell}(t)}{dt}$, aligning with the conventional description of self-torque in light from literature (7–10). This component of torque manifests as a change in orbital radius (since $\ell$ and $r$ are correlated), analogous to mechanical torque induced by a wrench.

The latter two terms arise from an azimuthal chirp in $\Delta\ell(\phi)$, leading to a variation in angular velocity. When a constant angular acceleration is assumed, the third term becomes negligible. These terms introduce a novel observation, suggesting an angular self-acceleration that occurs without an external torque. Unlike previous demonstrations, here no radius change occurs; instead, the change in OAM results from an increase in the spread of topological charge (i.e., $\Delta\ell$ increases) across the same frequency spectrum ($\Delta\omega$). This follows from the understanding that a polychromatic beam of light does not behave as a rigid body, as discussed in (3, 11), and as a result a chirp in $\Delta\ell$ leads to an energy redistribution across the frequencies, and thus an angular acceleration in order to conserve momentum.

## S7. Optimizing the range of achievable self-torque

The spectral resolution of our space-time shaper—the ability to separate and modulate frequencies with minimal crosstalk (i.e., the number of addressable frequency bands, $\Delta\ell + 1$)— emerges as a bottleneck in expanding the achievable range of self-torque. Currently, our system facilitates approximately 16 independently addressable frequency bands (refer to Methods). This, however, should be viewed as a limitation of our technical implementation, not the technique itself.

Additionally, the time-varying orbital angular momentum (OAM), which depends on the mean topological charge, offers potential for broadening the accessible range of self-torque. We have successfully demonstrated topological charges up to $\ell = 150$, surpassing earlier implementations of helical wave packets. Considering a typical annular beam radius of 4 mm in our system, with a SLM (HOLOEYE GAEA) pixel size of 3.74 μm and a requirement for at least 8 pixels per azimuthal sector to cover a phase ramp from 0 to $2\pi$, we estimate that the maximum topological charge achievable in our current configuration is approximately $\ell \approx 850$. However, this would necessitate consideration of the numerical aperture of the imaging system. To further extend the limits of our approach, exploring avenues such as employing shorter laser pulses with wider bandwidths or utilizing nano-structured devices like metasurfaces, known for their superior subwavelength resolution, could be promising strategies to achieve larger accelerations, reaching attosecond variations in OAM.

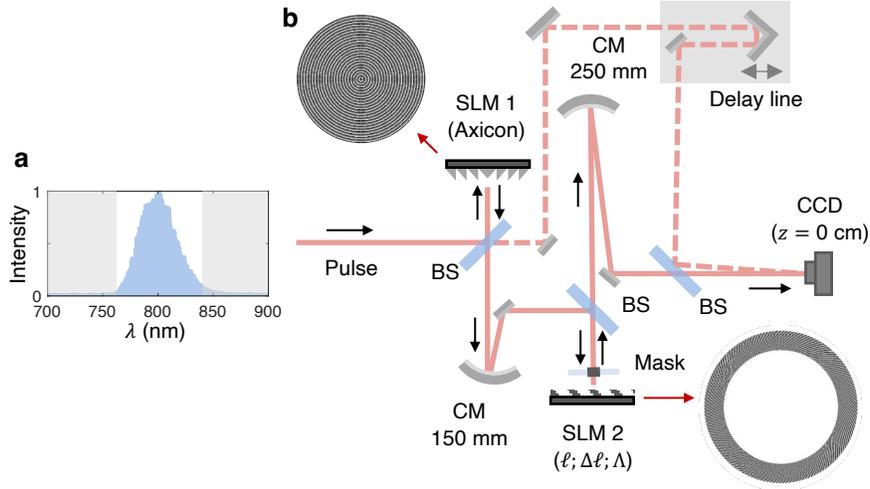

**Fig. S4. Digital Fourier space-time beam shaper for helical wave packets. a)** The intensity spectrum of the femtosecond pulses from the NOPA, centered at 800 nm and with a duration of 25 fs. The grey-shaded areas mask the unused part of the spectrum, giving access to a bandwidth of 80 nm. **b)** Schematic of the space-time shaper. The ultrashort pulse is incident normal an axicon grating, which is encoded as a phase hologram on the first spatial light modulator (SLM), which maps the spectral content into concentric rings in the far-field ($2f$). A metallic disk mask placed at $2f$ is used to block the zero order of the first SLM. A hologram displayed on a second SLM in the far field imparts a topological-spectral correlation. The helical wave packets are generated at the imaging plane of the first SLM ($4f$). The characterization setup combines a delay line and spatial interferometer based on off-axis digital holography. The interference images are recorded by the charge-coupled device camera located at the imaging plane ($4f$) of the first SLM. CM, concave mirror; BS, beamsplitter; CCD; charge-coupled device.

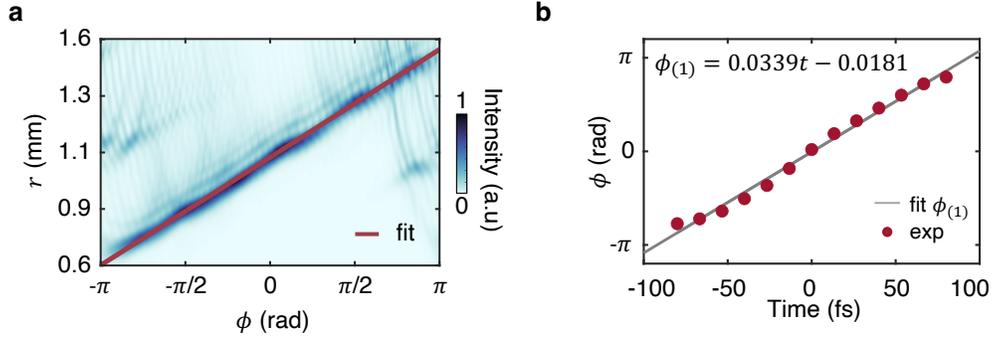

**Fig. S5. Additional angular dynamics of the time-varying topological charge helical wave packet. a**) The unwrapped intensity profile along the azimuthal coordinate of the experimentally generated helical wave packet in Fig. 2 of the main text ... orbit of the helical wave packet ... y with azimut ... of an Archimedes spiral. **b**) ... packet's angular ... ear fit confirms the expected ...

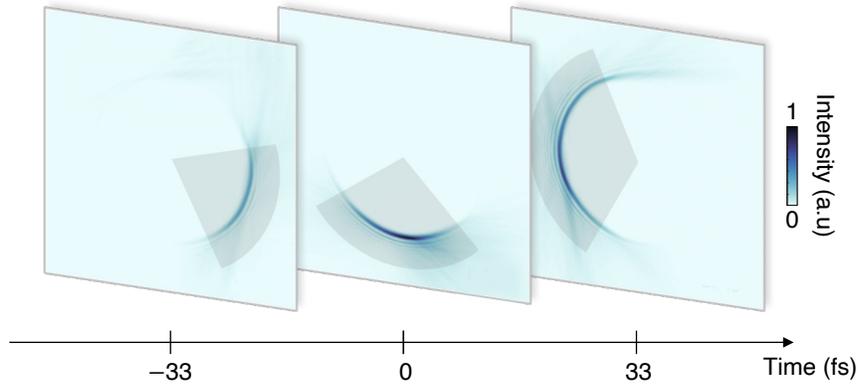

**Fig. S6. Energy redistribution in an angularly accelerating wave packet.** Intensity images of the experimentally generated helical wave packet in Fig. 3 of the main text, captured at three equal time intervals as denoted in Fig. 3d. The profiles are obtained from the wave packet's complex electric field, extracted via off-axis digital holography and a delay line. The sequence highlights the accelerated motion of the azimuthal wave packet, with the azimuthal position shifting noticeably. Furthermore, the arc subtended by the wave packet visibly increases over time, consistent with our interpretation that angular acceleration results from the redistribution of energy among wave packet components with a range of angular momentum values ($\Delta\ell$), across a constant bandwidth ($\Delta\omega$). The greater the $\Delta\ell$ spread, the more localized the angular position becomes.

**Movie S1.**

Movie illustrating the time evolution (at $z = 0$) of the wave packet reported in fig. S3, experimentally measured via off-axis holography. The movie shows: the intensity and phase profile at different time delays. The phase profile is cut at 10% max intensity. Notably, the wave packet reverses its orbital direction midway through the sequence. The window size is 4 mm by 4 mm.

**Movie S2.**

Movie illustrating the time evolution (at $z = 0$) of the wave packet reported in Fig. 2 of the main text, experimentally measured by off-axis holography. The movie shows: the intensity and phase profile at different time delays. The phase profile is cut at 10% max intensity. Notably, the wave packet spirals outward with increasing radius. The window size is 4 mm by 4 mm.

**Movie S3.**

Movie illustrating the time evolution (at $z = 0$) of the wave packet reported in Fig. 3 of the main text, experimentally measured via off-axis holography. The movie shows: the intensity and phase profile at different time delays. The phase profile is cut at 10% max intensity. Notably, the wave packet angularly self-accelerates. The window size is 4 mm by 4 mm.